\documentclass[prl,twocolumn,showpacs,amsmath,amssymb,amsfonts,superscriptaddress]{revtex4}
\usepackage{mathptmx}
\usepackage{epsfig}
\usepackage{dcolumn}
\usepackage{bm}

\begin{document}


\title{A Surface Reconstruction with a Fractional Hole: $(\sqrt{5}\times\sqrt{5}) R26.6^\circ$ LaAlO$_3$ (001)}
\author{Courtney H Lanier}%
	\thanks{The authors contributed equally to this work.}%
\author{James M Rondinelli}%
	\thanks{The authors contributed equally to this work.}%
	\altaffiliation[Now at: ]{Materials Department, University of California, Santa Barbara, CA, 93106-5050, USA}%
\author{Bin Deng}	
\author{Roar Kilaas}
	\affiliation{%
	Department of Materials Science and Engineering, Northwestern University, Evanston, Illinois, 60208, USA}%
\author{Kenneth R Poeppelmeier}
	\affiliation{%
	Department of Chemistry, Northwestern University, Evanston, Illinois 60208, USA}%
\author{Laurence D Marks}
 \email[Address correspondence to: ]{L-marks@northwestern.edu}
	\affiliation{%
	Department of Materials Science and Engineering, Northwestern University, Evanston, Illinois, 60208, USA}%
\date{\today}
\begin{abstract}
The structure of the $(\sqrt{5}\times\sqrt{5})R26.6^\circ$ reconstruction of LaAlO$_3$ (001) has been determined using transmission electron diffraction combined with direct methods.  The structure is relatively simple, consisting of a lanthanum oxide termination with one lanthanum cation vacancy per surface unit cell.  The electronic structure is unusual since a fractional number of holes or atomic occupancies per surface unit cell are required to achieve charge neutrality. Density functional calculations indicate that the charge compensation mechanism occurs by means of highly delocalized holes.  The surface contains no oxygen vacancies and with a better than 99\% confidence level, the holes are not filled with hydrogen.  The reconstruction can be understood in terms of expulsion of the more electropositive cation from the surface followed by an increased covalency between the remaining surface lanthanum atoms and adjacent oxygen atoms.  
\end{abstract}
\maketitle
Lanthanum aluminate (LaAlO$_3$, LAO) is an important oxide material as a substrate for thin film growth \cite{REF:Bro90,REF:Sch2001,REF:Sim88,REF:Sch2006}, a potential gate dielectric (or buffer layer) \cite{REF:Cab97,REF:For2005,REF:Hwa2006,REF:Kle2005,REF:Nak2006,REF:Par2001,REF:Oht2004}, as well as a catalyst \cite{REF:Yam2001,REF:Val96,REF:Liu99}. This transition metal oxide is representative of a larger class of materials with ABO$_3$ stoichiometry known as perovskites. There has recently been interest in the character of the interfaces between LaAlO$_3$ (001) and other materials such as Si and SrTiO$_3$ \cite{REF:Oht2002,REF:Kni2005}. Since both La and Al are almost always in a trivalent state, conventional methods of counting carriers for charge compensation lead to the unusual conclusion that a fractional number of excess carriers per unit cell is required. The effect of such correlated electron properties 
in oxide heterostructures 
has been studied extensively theoretically \cite{REF:Oka2004b,REF:Oka2005,REF:Fre2004,REF:Pop2005,REF:Oka2006} for these interfaces, since charge redistribution 
is well-described by the electrochemical potential gradient between the materials.

LaAlO$_3$ (001) consists of alternating layers of LaO and AlO$_2$ stacked along the $<$001$>$ cubic direction. Consequently, the formal charges of La$^{3+}$, Al$^{3+}$ and O$^{2-}$ produce two terminations differing in nominal charges of (La-O)$^{+}$ and (Al-O$_2$)$^{-}$, and an excess half electron (or hole) exists per unit interface cell. The LAO (001) surface is therefore polar and classified as a Type III surface within Tasker's convention \cite{REF:Tas79}. The composition, structure and morphology of LAO (001) has been researched in the past few years; however, structural information of how the surface terminates is still ambiguous. The majority of the analyses \cite{REF:Sch2006,REF:Fra2001,REF:Yao98,REF:Kaw2002,REF:Wan95a,REF:van98,REF:Jac97a,REF:Jac97b} have considered a (1$\times$1) surface prepared by relatively low-temperature annealing (the melting point of LaAlO$_3$ is 2110$^\circ$C) in various atmospheres, and report either (or both) LaO and AlO$_2$ terminations; it is not clear whether the surfaces have reached thermodynamic equilibrium. The only reported reconstructions on the (001) surface of LAO is a (5$\times$5) reconstruction obtained by annealing at 1500$^\circ$C for 20 hours in air \cite{REF:Wan95b}.  

To date, all models have failed to offer any feasible stabilization pathways that are consistent with the experimental results. Furthermore, very little of the work has addressed the issue of charge compensation and how to reconcile the dipole at the surface. In this Letter, we report experimental results for the  {$(\sqrt{5}\times\sqrt{5})R26.6^\circ$} surface reconstruction, hereafter referred to as \textsc{Rt5}, on the LaAlO$_3$(001) surface. We also provide a first-principles investigation of the structure and show how the polar surface is passivated through a redistribution of the near surface electron density.

Single crystal LaAlO$_3$ (001) wafers from MTI Corporation (99.95\% pure) were prepared for transmission electron microscopy studies using conventional methods of dimpling and ion-beam thinning (a Gatan Precision Ion Polishing System with 4.8~kV Ar$^{+}$ ions) until electron transparent. They were then annealed at temperatures between 1100-1500$^\circ$C in a Carbolite STF 15/51/180 tube furnace for three hours. While initial experiments were performed in air, we also annealed in a mix of 20\%~O$_2$~:~80\%~N$_2$, which mass-spectrometer measurements indicated had a maximum impurity level of 10~ppb of H$_2$O. 

Diffraction experiments were performed with a Hitachi 8100 electron microscope operating at 200~kV. Negatives with exposure times ranging from 0.5~to~90~s were recorded then digitized to 8 bits with a 25~$\mu$m pixel size on an Optronics P-1000 microdensitometer which was calibrated to be linear over the selected exposure range. Intensities measurements were determined using a cross-correlation technique \cite{REF:Xu94}, and the data sets (7911 surface reflections) were symmetry reduced using \emph{p4} plane group symmetry to 94 independent beams with the error for each reflection determined using robust statistical methods. These were analyzed using electron direct methods (\textsc{EDM 2.0.1}) software \cite{REF:Kil2006} and in-plane atomic positions refined against the experimental data gave a $\chi^{2}=4.83$.

For the theoretical modeling, the LAO surface structure was geometry optimized using a three-dimensional periodic DFT surface slab model of 9 layers (118 atoms) separated by 8~\AA~of vacuum. To analyze the charge density, calculations were performed using the all-electron (linearized) augmented-plane wave + local orbitals (L/APW+lo) method as implemented in WIEN2k \cite{REF:Bla2001} with the PBE-GGA \cite{REF:Per96b} exchange-correlation functional, a plane-wave cutoff of RK$_{\rm max}$=6.75 and muffin-tin radii of 1.75, 1,75 and 2.33 Bohr for O, Al and La respectively. To test for water splitting and oxygen vacancies, calculations were also performed using the projector augmented wave (PAW) approach \cite{REF:Blo94} as implemented in the Vienna \emph{Ab Initio} Simulation Package (VASP) code \cite{REF:Kre96,REF:Kre99} using a 3$\times$3$\times$1 k-point grid,  plane wave energy cutoff of 360~eV, electronic iteration convergence of 0.001~eV and geometry relaxation convergence of 0.01~eV. The Fermi surface was smeared using a Gaussian width of 0.20~eV. For these calculations we used the conventional LDA energy functional as well as the PBE \cite{REF:Per96a} and PW92 \cite{REF:Per92b} functionals. In all cases the surface unit cell used was matched to the DFT minimized lattice parameter for the appropriate functionals.

\begin{figure}
	\centering
		\includegraphics[width=0.48\textwidth]{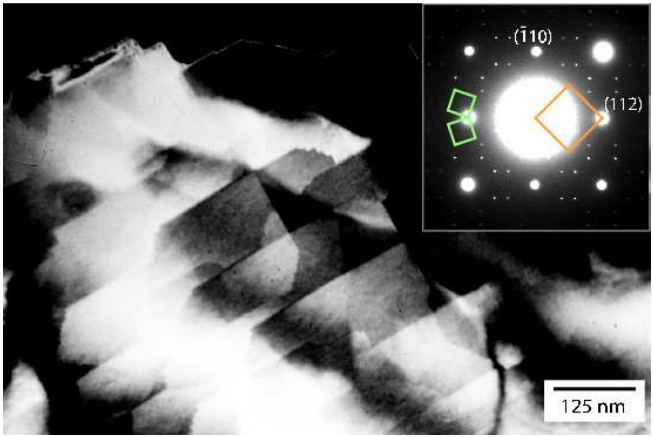}
		\caption{\label{fig:diffpattern}(color online).\ Dark field image showing extended $<$100$>$ faceting with step bunches and reconstructed terraces. A small probe off-zone diffraction pattern (inset) is shown with the rhombohedral (1$\times$1) bulk unit cell (orange, right) and the surface unit cell for both domains of the reconstruction (green, left).}
\end{figure}

\begin{figure}
	\centering
		\includegraphics[width=0.48\textwidth]{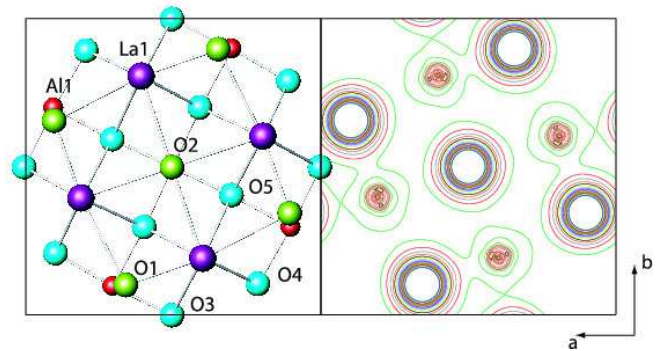}
		\caption{\label{fig:structure}(color online).\ Top view of the \textsc{Rt5} surface reconstruction (left) with the surface unit cell outlined in black (a=8.537 \AA). Contour map of the valence density in the surface plane (right) at $0.1~e$/\AA$^{3}$ showing the increased electron density between the surface oxygen (O1 and O2) and lanthanum atoms (La1) corresponding to increased covalent bonding. Atoms O1, O2 and La1 are in the surface layer, while Al1, O3, O4 and O5 are in the second layer.}
\end{figure}

A typical transmission electron diffraction pattern of the reconstructed surface is shown in Fig.~\ref{fig:diffpattern}. The surface is well ordered with facets evident in the dark field image and minimal diffuse background scattering from disorder present in the diffraction pattern (inset). Direct methods analysis and refinement were straightforward, and gave the structure shown in Fig.~\ref{fig:structure}; information about the atomic positions can be found in Table \ref{tab:positions}, while the experimental and calculated intensities are given in the supplemental material.

Bulk LAO has a rhombohedral to cubic phase transition at 435$\pm$25$^\circ$C.  Consequently, the \textsc{Rt5} reconstruction forms on a cubic LAO substrate, but this substrate becomes rhombohedral upon cooling.  As the structures of rhombohedral LAO and cubic LAO only differ by tenths of a degree, it is reasonable to expect the cubic surface reconstruction to be accommodated on bulk rhombohedral LAO at room temperature.  Nonetheless, the possibility of a rhombohedral surface structure (\emph{p2} plane symmetry) and a cubic surface structure (\emph{p4}, \emph{p4mm}, \emph{p4mg} plane symmetry) were both explored and a  \emph{p4} symmetry provided the best fit to the data.  The possibility of an aluminum overlayer was also considered, however the fit to the data was noticeably inferior.

The \textsc{Rt5} reconstruction is an overlayer of lanthanum oxide on the AlO$_2$ bulk termination of LAO with one lanthanum vacancy per surface unit cell. The surface stoichiometry may be written as $({V{\text{La}}_{\text{4}} {\text{O}}_{\text{5}}})^{-\frac{1}{2}}$, where $V$ is the lanthanum cation vacancy and the surface unit has a nominal charge of $-\frac{1}{2}$. Each surface lanthanum (La1) is coordinated to four oxygens within the surface layer and four oxygens in the layer below. Additionally, the lanthanum atom is displaced into the bulk by $\approx$0.20\AA, while the oxygen atoms are displaced away from the bulk. The oxygen atoms (O1 and O2) in the surface layer preserve the octahedral coordination of the aluminum atoms in the layer below; however due to the La vacancy, O1 is now only four-coordinate while O2 is five-coordinate.

The formation of the \textsc{Rt5} reconstruction can be understood as follows.  To reduce positive charge on the surface of an ideal La-O bulk termination, either La$^{3+}$ or Al$^{3+}$ cation vacancies are formed.  Because La$^{3+}$ is less electronegative than Al$^{3+}$, lanthaum cation vacancies are favored.  It follows that the surface bond covalency should increase to satisfy the under-coordination of the surface oxygens.  Therefore since the La-O bonds are longer compared to the Al-O bonds, and the non-bonding oxygen repulsive interactions are smaller than those of the Al-O octahedra, the La-O bonds become shorter, with an average La-O bond length of 2.61~\AA~(compared to 2.68~\AA~in the bulk).  However, despite these significant changes, a persistent, albeit reduced, polarization remains at the surface.


The lanthanum cation expulsion nearly reconciles the charge neutrality problem at the surface. For the \textsc{Rt5} reconstruction with an area equal to five interface unit cells of the bulk, it is impossible to form a fully charge compensated surface without invoking a fractional density (per reconstructed surface cell) of carriers or a partial occupancy of cation sites. While there was too little diffuse intensity in the diffraction data to support a fractional occupancy of sites, a low density of oxygen vacancies is undetectable experimentally. Similarly there is the possibility of disordered hydrogen atoms on the surface and several studies indicate that some bulk oxides (and surfaces) can contain low levels of hydrogen (\cite{REF:Wan2000,REF:Laz2005,REF:Nog2000,REF:Nog2004,REF:Wol2004,REF:Van2000}). 

To test for the presence of hydrogen on the surface, we considered the possible reaction
\[
4\left({{\textsc{Rt}}5} \right) + {\text{H}}_{\text{2}} {\text{O}} \rightleftharpoons 4\left({\textsc{Rt}}5{\text{H}}_{1/2} \right) + \tfrac{1}{2}{\text{O}}_2 
\]
with partial occupancy of a hydrogen bonded to O1 (lower in energy than the alternative O2). Four calculations were performed using VASP: one without hydrogen for a $\sqrt{10}\times\sqrt{10}$ supercell (9-layers, 236 atoms) rotated by $45^\circ$ containing two surfaces each with four \textsc{Rt}5 unit cells, another of the same cell but with $\frac{1}{2}$ hydrogen per \textsc{Rt5} surface unit cell  (atomic positions are given in the supplemental material) and two for the isolated molecules H$_2$O and O$_2$.
\begingroup
\squeezetable
\begin{table}
\caption{Atom positions for the \textsc{Rt5} reconstruction are given as fractional coordinates of the surface cell (${\rm a}=8.526$ \AA). The La vacancy is denoted by $V$ at the surface (layer 1) and the bulk corresponds to layer 5. $\Delta_{\rm z}=\left|z_{\rm DFT}-z_{\rm Bulk} \right|$ in \AA~and positive (negative) deviations indicate a displacement away from (into) the bulk. The excess charge values for each atom are denoted by $\delta$Q ($10^{-2}$~$e$), and the integrated hole density as $\rho(h)$ ($10^{-2}$~$e$/\AA$^{2}$). Bulk LAO provides the charge reference: ${\rm O}=-1.540e$, ${\rm La}=2.065e$ and ${\rm Al}=2.554e$.}
\label{tab:positions}
\begin{ruledtabular}
\begin{tabular}{clD{.}{.}{1.2}D{.}{.}{1.2}D{.}{.}{1.3}D{.}{.}{1.3}D{.}{.}{1.3}D{.}{.}{2.3}D{.}{.}{2.1}D{.}{.}{1.1}}
& & \multicolumn{2}{c}{Experiment}  & \multicolumn{3}{c}{\textsc{WIEN2k}} & & \\
\cline{3-7}
Layer & Atom &\multicolumn{1}{c}{$x$} &\multicolumn{1}{c}{$y$} &\multicolumn{1}{c}{$x$} &\multicolumn{1}{c}{$y$} & \multicolumn{1}{c}{$z$} & \multicolumn{1}{c}{$\Delta_{\rm z}$} &  \multicolumn{1}{c}{$\delta$Q} &  \multicolumn{1}{c}{$\rho(h)$}\\
\hline
1 & $V$ & 0.00&	0.00&	0.000	& 0.000 &	0.320	&\multicolumn{1}{c}{---}&\multicolumn{1}{c}{---}&\multicolumn{1}{c}{---}\\
	&	O1	&	0.34&	0.90& 0.344	&	0.898	&	0.320	&+0.036						&10.2	 & 2.1				\\
	&	O2	& 0.50&	0.50& 0.500	&	0.500	&	0.321	&+0.055						&5.8	 & 2.5				\\
	&	La1	& 0.81&	0.40& 0.808	& 0.396	& 0.310	&-0.196						&-2.6	 & 0.3				\\
2 & Al1	& 0.30&	0.90& 0.298	& 0.901	& 0.240	&+0.035						&-0.7	 & 0.0				\\
	&	Al2	& 0.50&	0.50& 0.500	& 0.500	& 0.239	&+0.015						&-0.7	 & 0.0				\\
	&	O3	& 0.50&	0.00& 0.500	&	0.000	& 0.232	&-0.169						&4.1	 & 2.0				\\
	&	O4	& 0.21&	0.11& 0.210	& 0.106	& 0.239	&+0.009						&1.4	 & 0.6				\\
	&	O5	& 0.70&	0.59& 0.702	& 0.594	& 0.237	&-0.041						&2.0	 & 1.3				\\
3 &	O6	& 0.29& 0.90& 0.290	& 0.899	& 0.159	&-0.007						&2.0	 & 1.8				\\
	&	O7	& 0.50& 0.50& 0.500	& 0.500	& 0.158	&-0.018						&2.2	 & 2.5				\\
	&	La2 & 0.80& 0.40& 0.800	& 0.399	& 0.158	&-0.036						&-0.3	 & 0.2				\\
	&	La3 & 0.00& 0.00& 0.000	& 0.000	& 0.163	&+0.099						&1.9	 & 0.2				\\
4 &	Al3 & 0.30& 0.90& 0.298	& 0.900	& 0.079	&-0.003						&0.0	 & 0.0				\\
	&	Al4 & 0.50& 0.50&	0.500	& 0.500	& 0.079	&-0.008						&0.3	 & 0.0				\\
	&	O8	& 0.50& 0.00& 0.500	& 0.000	& 0.081	&+0.038				    &0.7	 & 0.7	 			\\
	&	O9	& 0.20& 0.10&	0.201	& 0.100	& 0.078	&-0.045						&0.4	 & 1.3				\\
	&	O10	& 0.70& 0.60& 0.700	& 0.600	& 0.079	&-0.009						&0.5	 & 1.0				\\
5	&	O11	& 0.29& 0.90& 0.304	& 0.900	& 0.000 &\multicolumn{1}{c}{n/a} &1.3 & 1.9				\\
	&	O12 & 0.50& 0.50& 0.500	& 0.500 & 0.000 &\multicolumn{1}{c}{n/a} &0.4 & 1.0				\\
	&	La4 & 0.80& 0.40& 0.800	& 0.400	& 0.000 &\multicolumn{1}{c}{n/a} &-0.2& 0.2				\\
	&	La5 & 0.00& 0.00& 0.000	& 0.000	& 0.000 &\multicolumn{1}{c}{n/a} &0.3 & 0.2				\\
\end{tabular}
\end{ruledtabular}
\end{table}
\endgroup
The DFT calculations indicated that this reaction can occur with an energy change of -0.99~eV at T=0~K for the PBE functional, -1.04~eV with the PW92 functional and -1.46~eV for the LDA. By comparing the PBE and PW92 numbers we can estimate that the intrinsic surface uncertainty \cite{REF:Mat2006} is small, about  0.15~eV. A reasonable estimate of the error in the energies is ${\left|E_{\rm LDA}-E_{\rm PBE}\right|}/{2}$, or 0.24~eV. Thus, at a 99\% confidence level (3$\sigma$) the maximum energy for this exothermic reaction would be $-1.71$~eV at T=0~K. Using tabulated values for the free-energy of water and oxygen \cite{refdata} the reaction becomes endothermic and would require more than 3.64~eV to take place at the \textsc{Rt5} formation temperature of 1200$^\circ$C  in an air atmosphere with less than 10~ppb water impurity. Hence, to a better than 99\% confidence level when the reconstruction is formed at high temperature, it is energetically unfavorable for the surface to split water and incorporate H$^{+}$. On cooling and exposure to environmental humidity it is quite possible that the surface splits water \cite{REF:STP}.
\begin{figure}
	\centering
		\includegraphics[width=0.48\textwidth]{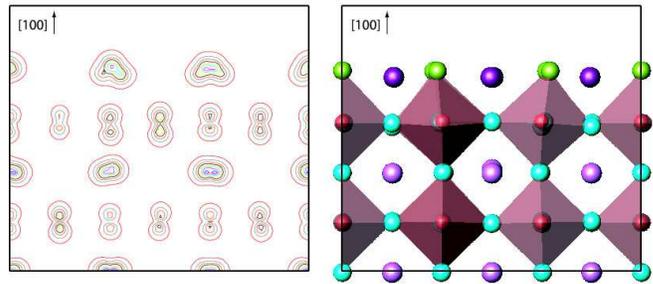}
		\caption{\label{fig:density1}(color online).\ Integrated hole density plot along the [100] direction at $0.01~e$/\AA$^{2}$ shown next to the crystal structure.}
\end{figure}  

Similarly, to test for the presence of oxygen vacancies on the \textsc{Rt5} surface, we considered two reduced structures with a $\frac{1}{4}$ oxygen vacancy per unit cell at sites O1 and O2 (the structures are available in the supplemental material). It was found that at T=0~K the lower energy reduced structure (vacancy at O1) is 2.58~eV higher in energy than the \textsc{Rt5} structure as presented. At the reconstruction formation temperature, the oxygen chemical potential is large enough to reduce the surface with an energy gain of -0.29~eV; however within the error associated with these computational methods, we posit the probability for this reduction is small. Furthermore, as the sample is cooled the oxygen vacancies (if any) would be filled and the \textsc{Rt5} surface is preserved. Again, we find that an alternative charge compensation mechanism is required, and suggest that an electron hole fulfills this requirement.

Hole densities were calculated  by integrating over the Bader volumes \cite{REF:Holes}. Somewhat unusual is that the hole densities do not decay off into the bulk, but rather the fractional hole is very delocalized over all the oxygen atoms, as indicated by the projected hole density of states along the [100] direction normal to the surface shown in Fig.~\ref{fig:density1}. This behavior is atypical for most bulk oxides since hole trapping occurs through acceptor defect sites in open lattice perovskite structures \cite{Donnerberg}. This effect leads to small hole mobilities in bulk materials, and is reminiscent of the  Mott-Wannier exciton model in more ionic crystals (despite the increased covalency here).

Although the charge on the atoms in a solid cannot be uniquely partitioned, there are several theoretical models which allow for its estimation; we used Bader's \emph{atom-in-molecule} (AIM) method \cite{bader} with the WIEN2k densities. Table \ref{tab:positions} includes the excess charges ($\delta \rm Q$) calculated for the various layers. There is a noticeable decrease in the charge on the surface oxygen atoms (O1 and O2) as well as the surface lanthanum (La1), with smaller variations decaying off more rapidly into the bulk for the other oxygen atoms. In bulk LaAlO$_3$ the Al-O bond has more covalent character than the La-O bond; the charge densities at the (3,-1) bond critical points, which are typically used as a qualitative estimate, are $\approx0.45$ and 0.24~eV/\AA$^3$ for the two.  At the surface there is essentially no change in the covalent character of the Al-O bonds, but the density between the La-O at the surface increases dramatically to a value of $\approx0.50$~eV/\AA$^3$~for the closest oxygen atom which is only 2.37~\AA~away compared to 2.70~\AA~in the bulk (with the DFT optimized lattice parameters). This suggests that some of the charge on the oxygen ions in the bulk transfers into the covalent bonds near the surface, and rather independent of this, it occurs in the presence of a highly delocalized hole.

For the \textsc{Rt5} reconstruction on the LaAlO$_3$ (001) surface, the surface polarity is quenched by the presence of the electron hole. This point has strong implications on the growth of nano-layer composites and the dynamics describing surface reconstructions, that is, interface passivation is highly sensitive to the experimental preparation. Furthermore, it is likely that oxygen vacancies can strongly influence the formation of charge carriers at such interfaces; while the results suggest that they are filled upon annealing, it is important to recognize that their electron donor character may behave differently at interfaces. We find that the driving force for reconstruction is the polar discontinuity at the surface and charge neutrality  requires either (or both) atomic or electronic reconfiguration: the expulsion of the lanthanum cation and delocalized electron hole meets this challenge.

\begin{acknowledgments}
This work was funded by NSF under Grant \#DMR-0455371 (JMR) and by the DOE under Contract \#DE-FG02-01ER45945/A006 (BD and RK) and \#DE-FG02-03ER15457 (CHL). JMR acknowledges support from the Office of the Provost at Northwestern University.
\end{acknowledgments}
\bibliographystyle{apsrev}
\bibliography{laoBIB}

\begin{thebibliography}{51}
\expandafter\ifx\csname natexlab\endcsname\relax\def\natexlab#1{#1}\fi
\expandafter\ifx\csname bibnamefont\endcsname\relax
  \def\bibnamefont#1{#1}\fi
\expandafter\ifx\csname bibfnamefont\endcsname\relax
  \def\bibfnamefont#1{#1}\fi
\expandafter\ifx\csname citenamefont\endcsname\relax
  \def\citenamefont#1{#1}\fi
\expandafter\ifx\csname url\endcsname\relax
  \def\url#1{\texttt{#1}}\fi
\expandafter\ifx\csname urlprefix\endcsname\relax\def\urlprefix{URL }\fi
\providecommand{\bibinfo}[2]{#2}
\providecommand{\eprint}[2][]{\url{#2}}

\bibitem[{\citenamefont{Brown~\emph{et al.}}(1990)}]{REF:Bro90}
\bibinfo{author}{\bibfnamefont{R.}~\bibnamefont{Brown~\emph{et al.}}},
  \bibinfo{journal}{Appl.\ Phys.\ Lett.} \textbf{\bibinfo{volume}{57}},
  \bibinfo{pages}{1351} (\bibinfo{year}{1990}).

\bibitem[{\citenamefont{Schneidewind et~al.}(2001)\citenamefont{Schneidewind,
  Manzel, and Kirsch}}]{REF:Sch2001}
\bibinfo{author}{\bibfnamefont{H.}~\bibnamefont{Schneidewind}},
  \bibinfo{author}{\bibfnamefont{M.}~\bibnamefont{Manzel}}, \bibnamefont{and}
  \bibinfo{author}{\bibfnamefont{K.}~\bibnamefont{Kirsch}},
  \bibinfo{journal}{Applied Superconductivity, IEEE Transactions on}
  \textbf{\bibinfo{volume}{11}}, \bibinfo{pages}{3106} (\bibinfo{year}{2001}).

\bibitem[{\citenamefont{Simon~\emph{et al.}}(1988)}]{REF:Sim88}
\bibinfo{author}{\bibfnamefont{R.~W.} \bibnamefont{Simon~\emph{et al.}}},
  \bibinfo{journal}{Appl.\ Phys.\ Lett.} \textbf{\bibinfo{volume}{53}},
  \bibinfo{pages}{2677} (\bibinfo{year}{1988}).

\bibitem[{\citenamefont{Schmidt~\emph{et al.}}(2006)}]{REF:Sch2006}
\bibinfo{author}{\bibfnamefont{D.~A.} \bibnamefont{Schmidt~\emph{et al.}}},
  \bibinfo{journal}{J.\ Appl.\ Phys.} \textbf{\bibinfo{volume}{99}}
  (\bibinfo{year}{2006}), \bibinfo{note}{113521}.

\bibitem[{\citenamefont{Cabanas~\emph{et al.}}(1997)}]{REF:Cab97}
\bibinfo{author}{\bibfnamefont{M.~V.} \bibnamefont{Cabanas~\emph{et al.}}},
  \bibinfo{journal}{Solid State Ionics} \textbf{\bibinfo{volume}{101}},
  \bibinfo{pages}{191} (\bibinfo{year}{1997}).

\bibitem[{\citenamefont{Forst et~al.}(2005)\citenamefont{Forst, Schwarz, and
  Blochl}}]{REF:For2005}
\bibinfo{author}{\bibfnamefont{C.~J.} \bibnamefont{Forst}},
  \bibinfo{author}{\bibfnamefont{K.}~\bibnamefont{Schwarz}}, \bibnamefont{and}
  \bibinfo{author}{\bibfnamefont{P.~E.} \bibnamefont{Blochl}},
  \bibinfo{journal}{Phys.\ Rev.\ Lett.} \textbf{\bibinfo{volume}{95}}
  (\bibinfo{year}{2005}), \bibinfo{note}{137602}.

\bibitem[{\citenamefont{Hwang}(2006)}]{REF:Hwa2006}
\bibinfo{author}{\bibfnamefont{H.~Y.} \bibnamefont{Hwang}},
  \bibinfo{journal}{MRS Bulletin} \textbf{\bibinfo{volume}{31}},
  \bibinfo{pages}{28} (\bibinfo{year}{2006}).

\bibitem[{\citenamefont{Klenov et~al.}(2005)\citenamefont{Klenov, Schlom, Li,
  and Stemmer}}]{REF:Kle2005}
\bibinfo{author}{\bibfnamefont{D.~O.} \bibnamefont{Klenov}},
  \bibinfo{author}{\bibfnamefont{D.~G.} \bibnamefont{Schlom}},
  \bibinfo{author}{\bibfnamefont{H.}~\bibnamefont{Li}}, \bibnamefont{and}
  \bibinfo{author}{\bibfnamefont{S.}~\bibnamefont{Stemmer}},
  \bibinfo{journal}{Jpn.\ J.\ Appl.\ Phys., Part 2}
  \textbf{\bibinfo{volume}{44}}, \bibinfo{pages}{L617} (\bibinfo{year}{2005}).

\bibitem[{\citenamefont{Nakagawa et~al.}(2006)\citenamefont{Nakagawa, Hwang,
  and Muller}}]{REF:Nak2006}
\bibinfo{author}{\bibfnamefont{N.}~\bibnamefont{Nakagawa}},
  \bibinfo{author}{\bibfnamefont{H.~Y.} \bibnamefont{Hwang}}, \bibnamefont{and}
  \bibinfo{author}{\bibfnamefont{D.~A.} \bibnamefont{Muller}},
  \bibinfo{journal}{Nat.\ Mater.} \textbf{\bibinfo{volume}{5}},
  \bibinfo{pages}{204} (\bibinfo{year}{2006}).

\bibitem[{\citenamefont{Park and Ishiwara}(2001)}]{REF:Par2001}
\bibinfo{author}{\bibfnamefont{B.-E.} \bibnamefont{Park}} \bibnamefont{and}
  \bibinfo{author}{\bibfnamefont{H.}~\bibnamefont{Ishiwara}},
  \bibinfo{journal}{Appl.\ Phys.\ Lett.} \textbf{\bibinfo{volume}{79}},
  \bibinfo{pages}{806} (\bibinfo{year}{2001}).

\bibitem[{\citenamefont{Ohtomo and Hwang}(2004)}]{REF:Oht2004}
\bibinfo{author}{\bibfnamefont{A.}~\bibnamefont{Ohtomo}} \bibnamefont{and}
  \bibinfo{author}{\bibfnamefont{H.~Y.} \bibnamefont{Hwang}},
  \bibinfo{journal}{Nature} \textbf{\bibinfo{volume}{427}},
  \bibinfo{pages}{423} (\bibinfo{year}{2004}).

\bibitem[{\citenamefont{Yamamoto~\emph{et al.}}(2001)}]{REF:Yam2001}
\bibinfo{author}{\bibfnamefont{T.}~\bibnamefont{Yamamoto~\emph{et al.}}},
  \bibinfo{journal}{J.\ Synch.\ Rad.} \textbf{\bibinfo{volume}{8}},
  \bibinfo{pages}{634} (\bibinfo{year}{2001}).

\bibitem[{\citenamefont{Valden~\emph{et al.}}(1996)}]{REF:Val96}
\bibinfo{author}{\bibfnamefont{M.}~\bibnamefont{Valden~\emph{et al.}}},
  \bibinfo{journal}{J.\ Catal.} \textbf{\bibinfo{volume}{161}},
  \bibinfo{pages}{614} (\bibinfo{year}{1996}).

\bibitem[{\citenamefont{Liu~\emph{et al.}}(1999)}]{REF:Liu99}
\bibinfo{author}{\bibfnamefont{S.}~\bibnamefont{Liu~\emph{et al.}}},
  \bibinfo{journal}{Catal.\ Lett.} \textbf{\bibinfo{volume}{63}},
  \bibinfo{pages}{167} (\bibinfo{year}{1999}).

\bibitem[{\citenamefont{Ohtomo~\emph{et al.}}(2002)}]{REF:Oht2002}
\bibinfo{author}{\bibfnamefont{A.}~\bibnamefont{Ohtomo~\emph{et al.}}},
  \bibinfo{journal}{Nature} \textbf{\bibinfo{volume}{419}},
  \bibinfo{pages}{378} (\bibinfo{year}{2002}).

\bibitem[{\citenamefont{Knizhnik~\emph{et al.}}(2005)}]{REF:Kni2005}
\bibinfo{author}{\bibfnamefont{A.~A.} \bibnamefont{Knizhnik~\emph{et al.}}},
  \bibinfo{journal}{Phys.\ Rev.\ B} \textbf{\bibinfo{volume}{72}},
  \bibinfo{pages}{235329} (\bibinfo{year}{2005}).

\bibitem[{\citenamefont{Okamoto and Millis}(2004)}]{REF:Oka2004b}
\bibinfo{author}{\bibfnamefont{S.}~\bibnamefont{Okamoto}} \bibnamefont{and}
  \bibinfo{author}{\bibfnamefont{A.~J.} \bibnamefont{Millis}},
  \bibinfo{journal}{Phys.\ Rev.\ B} \textbf{\bibinfo{volume}{70}},
  \bibinfo{pages}{241104(R)} (\bibinfo{year}{2004}).

\bibitem[{\citenamefont{Okamoto and Millis}(2005)}]{REF:Oka2005}
\bibinfo{author}{\bibfnamefont{S.}~\bibnamefont{Okamoto}} \bibnamefont{and}
  \bibinfo{author}{\bibfnamefont{A.~J.} \bibnamefont{Millis}},
  \bibinfo{journal}{Phys.\ Rev.\ B} \textbf{\bibinfo{volume}{72}},
  \bibinfo{pages}{235108} (\bibinfo{year}{2005}).

\bibitem[{\citenamefont{Freericks}(2004)}]{REF:Fre2004}
\bibinfo{author}{\bibfnamefont{J.~K.} \bibnamefont{Freericks}},
  \bibinfo{journal}{Phys.\ Rev.\ B} \textbf{\bibinfo{volume}{70}},
  \bibinfo{pages}{195342} (\bibinfo{year}{2004}).

\bibitem[{\citenamefont{Popovic and Satpathy}(2005)}]{REF:Pop2005}
\bibinfo{author}{\bibfnamefont{Z.~S.} \bibnamefont{Popovic}} \bibnamefont{and}
  \bibinfo{author}{\bibfnamefont{S.}~\bibnamefont{Satpathy}},
  \bibinfo{journal}{Phys.\ Rev.\ Lett.} \textbf{\bibinfo{volume}{94}},
  \bibinfo{pages}{176805} (\bibinfo{year}{2005}).

\bibitem[{\citenamefont{Okamoto et~al.}(2006)\citenamefont{Okamoto, Millis, and
  Spaldin}}]{REF:Oka2006}
\bibinfo{author}{\bibfnamefont{S.}~\bibnamefont{Okamoto}},
  \bibinfo{author}{\bibfnamefont{A.~J.} \bibnamefont{Millis}},
  \bibnamefont{and} \bibinfo{author}{\bibfnamefont{N.~A.}
  \bibnamefont{Spaldin}}, \bibinfo{journal}{Phys.\ Rev.\ Lett.}
  \textbf{\bibinfo{volume}{97}}, \bibinfo{pages}{056802}
  (\bibinfo{year}{2006}).

\bibitem[{\citenamefont{Tasker}(1979)}]{REF:Tas79}
\bibinfo{author}{\bibfnamefont{P.~W.} \bibnamefont{Tasker}},
  \bibinfo{journal}{Journal of Physics: Condensed Matter}
  \textbf{\bibinfo{volume}{12}}, \bibinfo{pages}{4977} (\bibinfo{year}{1979}).

\bibitem[{\citenamefont{Francis et~al.}(2001)\citenamefont{Francis, Moss, and
  Jacobson}}]{REF:Fra2001}
\bibinfo{author}{\bibfnamefont{R.~J.} \bibnamefont{Francis}},
  \bibinfo{author}{\bibfnamefont{S.~C.} \bibnamefont{Moss}}, \bibnamefont{and}
  \bibinfo{author}{\bibfnamefont{A.~J.} \bibnamefont{Jacobson}},
  \bibinfo{journal}{Phys.\ Rev.\ B} \textbf{\bibinfo{volume}{64}},
  \bibinfo{pages}{235425} (\bibinfo{year}{2001}).

\bibitem[{\citenamefont{Yao~\emph{et al.}}(1998)}]{REF:Yao98}
\bibinfo{author}{\bibfnamefont{J.}~\bibnamefont{Yao~\emph{et al.}}},
  \bibinfo{journal}{J.\ Chem.\ Phys.} \textbf{\bibinfo{volume}{108}},
  \bibinfo{pages}{1645} (\bibinfo{year}{1998}).

\bibitem[{\citenamefont{Kawanowa~\emph{et al.}}(2002)}]{REF:Kaw2002}
\bibinfo{author}{\bibfnamefont{H.}~\bibnamefont{Kawanowa~\emph{et al.}}},
  \bibinfo{journal}{Surf.\ Sci.} \textbf{\bibinfo{volume}{506}},
  \bibinfo{pages}{87} (\bibinfo{year}{2002}).

\bibitem[{\citenamefont{Wang and Shapiro}(1995{\natexlab{a}})}]{REF:Wan95a}
\bibinfo{author}{\bibfnamefont{Z.~L.} \bibnamefont{Wang}} \bibnamefont{and}
  \bibinfo{author}{\bibfnamefont{A.~J.} \bibnamefont{Shapiro}},
  \bibinfo{journal}{Surf.\ Sci.} \textbf{\bibinfo{volume}{328}},
  \bibinfo{pages}{141} (\bibinfo{year}{1995}{\natexlab{a}}).

\bibitem[{\citenamefont{van~der Heide and Rabalais}(1998)}]{REF:van98}
\bibinfo{author}{\bibfnamefont{P.~A.~W.} \bibnamefont{van~der Heide}}
  \bibnamefont{and} \bibinfo{author}{\bibfnamefont{J.~W.}
  \bibnamefont{Rabalais}}, \bibinfo{journal}{Chem.\ Phys.\ Lett.}
  \textbf{\bibinfo{volume}{297}}, \bibinfo{pages}{350} (\bibinfo{year}{1998}).

\bibitem[{\citenamefont{Jacobs et~al.}(1997)\citenamefont{Jacobs, Miguel, and
  Alvarez}}]{REF:Jac97a}
\bibinfo{author}{\bibfnamefont{J.-P.} \bibnamefont{Jacobs}},
  \bibinfo{author}{\bibfnamefont{M.~A.~S.} \bibnamefont{Miguel}},
  \bibnamefont{and} \bibinfo{author}{\bibfnamefont{L.~J.}
  \bibnamefont{Alvarez}}, \bibinfo{journal}{THEOCHEM}
  \textbf{\bibinfo{volume}{390}}, \bibinfo{pages}{193} (\bibinfo{year}{1997}).

\bibitem[{\citenamefont{Jacobs~\emph{et al.}}(1997)}]{REF:Jac97b}
\bibinfo{author}{\bibfnamefont{J.-P.} \bibnamefont{Jacobs~\emph{et al.}}},
  \bibinfo{journal}{Surf.\ Sci.} \textbf{\bibinfo{volume}{389}},
  \bibinfo{pages}{L1147} (\bibinfo{year}{1997}).

\bibitem[{\citenamefont{Wang and Shapiro}(1995{\natexlab{b}})}]{REF:Wan95b}
\bibinfo{author}{\bibfnamefont{Z.~L.} \bibnamefont{Wang}} \bibnamefont{and}
  \bibinfo{author}{\bibfnamefont{A.~J.} \bibnamefont{Shapiro}},
  \bibinfo{journal}{Surf.\ Sci.} \textbf{\bibinfo{volume}{328}},
  \bibinfo{pages}{159} (\bibinfo{year}{1995}{\natexlab{b}}).

\bibitem[{\citenamefont{Xu et~al.}(1994)\citenamefont{Xu, Jayaram, and
  Marks}}]{REF:Xu94}
\bibinfo{author}{\bibfnamefont{P.}~\bibnamefont{Xu}},
  \bibinfo{author}{\bibfnamefont{G.}~\bibnamefont{Jayaram}}, \bibnamefont{and}
  \bibinfo{author}{\bibfnamefont{L.~D.} \bibnamefont{Marks}},
  \bibinfo{journal}{Ultramicroscopy} \textbf{\bibinfo{volume}{53}},
  \bibinfo{pages}{15} (\bibinfo{year}{1994}).

\bibitem[{\citenamefont{Kilaas~\emph{et al.}}(2006)}]{REF:Kil2006}
\bibinfo{author}{\bibfnamefont{R.}~\bibnamefont{Kilaas~\emph{et al.}}},
  \emph{\bibinfo{title}{Edm: Electron direct methods, v2.0.1}}
  (\bibinfo{year}{2006}), \urlprefix\url{www.numis.northwestern.edu/edm}.

\bibitem[{\citenamefont{Blaha~\emph{et al.}}(2001)}]{REF:Bla2001}
\bibinfo{author}{\bibfnamefont{P.}~\bibnamefont{Blaha~\emph{et al.}}},
  \emph{\bibinfo{title}{An Augmented Plane Wave + Local Orbitals Program for
  Calculating Crystal Properties}} (\bibinfo{publisher}{Karlheinz Schwarz,
  Techn. Universitat Wien, Austria}, \bibinfo{year}{2001}).

\bibitem[{\citenamefont{Perdew et~al.}(1996{\natexlab{a}})\citenamefont{Perdew,
  Burke, and Wang}}]{REF:Per96b}
\bibinfo{author}{\bibfnamefont{J.~P.} \bibnamefont{Perdew}},
  \bibinfo{author}{\bibfnamefont{K.}~\bibnamefont{Burke}}, \bibnamefont{and}
  \bibinfo{author}{\bibfnamefont{Y.}~\bibnamefont{Wang}},
  \bibinfo{journal}{Phys.\ Rev.\ B} \textbf{\bibinfo{volume}{54}},
  \bibinfo{pages}{16533} (\bibinfo{year}{1996}{\natexlab{a}}).

\bibitem[{\citenamefont{Blochl}(1994)}]{REF:Blo94}
\bibinfo{author}{\bibfnamefont{P.~E.} \bibnamefont{Blochl}},
  \bibinfo{journal}{Phys.\ Rev.\ B} \textbf{\bibinfo{volume}{50}},
  \bibinfo{pages}{17953} (\bibinfo{year}{1994}).

\bibitem[{\citenamefont{Kresse and Furthmuller}(1996)}]{REF:Kre96}
\bibinfo{author}{\bibfnamefont{G.}~\bibnamefont{Kresse}} \bibnamefont{and}
  \bibinfo{author}{\bibfnamefont{J.}~\bibnamefont{Furthmuller}},
  \bibinfo{journal}{Phys.\ Rev.\ B} \textbf{\bibinfo{volume}{54}},
  \bibinfo{pages}{11169} (\bibinfo{year}{1996}).

\bibitem[{\citenamefont{Kresse and Joubert}(1999)}]{REF:Kre99}
\bibinfo{author}{\bibfnamefont{G.}~\bibnamefont{Kresse}} \bibnamefont{and}
  \bibinfo{author}{\bibfnamefont{D.}~\bibnamefont{Joubert}},
  \bibinfo{journal}{Phys.\ Rev.\ B} \textbf{\bibinfo{volume}{59}},
  \bibinfo{pages}{1758} (\bibinfo{year}{1999}).

\bibitem[{\citenamefont{Perdew et~al.}(1996{\natexlab{b}})\citenamefont{Perdew,
  Burke, and Ernzerhof}}]{REF:Per96a}
\bibinfo{author}{\bibfnamefont{J.~P.} \bibnamefont{Perdew}},
  \bibinfo{author}{\bibfnamefont{K.}~\bibnamefont{Burke}}, \bibnamefont{and}
  \bibinfo{author}{\bibfnamefont{M.}~\bibnamefont{Ernzerhof}},
  \bibinfo{journal}{Phys.\ Rev.\ Lett.} \textbf{\bibinfo{volume}{77}},
  \bibinfo{pages}{3865} (\bibinfo{year}{1996}{\natexlab{b}}).

\bibitem[{\citenamefont{Perdew and Wang}(1992)}]{REF:Per92b}
\bibinfo{author}{\bibfnamefont{J.~P.} \bibnamefont{Perdew}} \bibnamefont{and}
  \bibinfo{author}{\bibfnamefont{Y.}~\bibnamefont{Wang}},
  \bibinfo{journal}{Phys.\ Rev.\ B} \textbf{\bibinfo{volume}{45}},
  \bibinfo{pages}{13244} (\bibinfo{year}{1992}).

\bibitem[{\citenamefont{Wang et~al.}(2000)\citenamefont{Wang, Chaka, and
  Scheffler}}]{REF:Wan2000}
\bibinfo{author}{\bibfnamefont{X.-G.} \bibnamefont{Wang}},
  \bibinfo{author}{\bibfnamefont{A.}~\bibnamefont{Chaka}}, \bibnamefont{and}
  \bibinfo{author}{\bibfnamefont{M.}~\bibnamefont{Scheffler}},
  \bibinfo{journal}{Phys.\ Rev.\ Lett.} \textbf{\bibinfo{volume}{84}},
  \bibinfo{pages}{3650} (\bibinfo{year}{2000}).

\bibitem[{\citenamefont{Lazarov~\emph{et al.}}(2005)}]{REF:Laz2005}
\bibinfo{author}{\bibfnamefont{V.~K.} \bibnamefont{Lazarov~\emph{et al.}}},
  \bibinfo{journal}{Phys.\ Rev.\ B} \textbf{\bibinfo{volume}{71}},
  \bibinfo{pages}{115434} (\bibinfo{year}{2005}).

\bibitem[{\citenamefont{Noguera}(2000)}]{REF:Nog2000}
\bibinfo{author}{\bibfnamefont{C.}~\bibnamefont{Noguera}},
  \bibinfo{journal}{Journal of Physics: Condensed Matter}
  \textbf{\bibinfo{volume}{12}}, \bibinfo{pages}{R367} (\bibinfo{year}{2000}).

\bibitem[{\citenamefont{Noguera et~al.}(2004)\citenamefont{Noguera, Finocchi,
  and Goniakowski}}]{REF:Nog2004}
\bibinfo{author}{\bibfnamefont{C.}~\bibnamefont{Noguera}},
  \bibinfo{author}{\bibfnamefont{F.}~\bibnamefont{Finocchi}}, \bibnamefont{and}
  \bibinfo{author}{\bibfnamefont{J.}~\bibnamefont{Goniakowski}},
  \bibinfo{journal}{Journal of Physics: Condensed Matter}
  \textbf{\bibinfo{volume}{16}}, \bibinfo{pages}{S2509} (\bibinfo{year}{2004}).

\bibitem[{\citenamefont{Woll}(2004)}]{REF:Wol2004}
\bibinfo{author}{\bibfnamefont{C.}~\bibnamefont{Woll}},
  \bibinfo{journal}{Journal of Physics: Condensed Matter}
  \textbf{\bibinfo{volume}{16}}, \bibinfo{pages}{S2981} (\bibinfo{year}{2004}).

\bibitem[{\citenamefont{Van~de Walle}(2000)}]{REF:Van2000}
\bibinfo{author}{\bibfnamefont{C.~G.} \bibnamefont{Van~de Walle}},
  \bibinfo{journal}{Phys.\ Rev.\ Lett.} \textbf{\bibinfo{volume}{85}},
  \bibinfo{pages}{1012} (\bibinfo{year}{2000}).

\bibitem[{\citenamefont{Mattsson et~al.}(2006)\citenamefont{Mattsson, Armiento,
  Schultz, and Mattsson}}]{REF:Mat2006}
\bibinfo{author}{\bibfnamefont{A.~E.} \bibnamefont{Mattsson}},
  \bibinfo{author}{\bibfnamefont{R.}~\bibnamefont{Armiento}},
  \bibinfo{author}{\bibfnamefont{P.~A.} \bibnamefont{Schultz}},
  \bibnamefont{and} \bibinfo{author}{\bibfnamefont{T.~R.}
  \bibnamefont{Mattsson}}, \bibinfo{journal}{Phys.\ Rev.\ B}
  \textbf{\bibinfo{volume}{73}}, \bibinfo{pages}{195123}
  (\bibinfo{year}{2006}).

\bibitem[{\citenamefont{{M. W. Chase Jr. \emph{et al.}}}(1985)}]{refdata}
\bibinfo{author}{\bibnamefont{{M. W. Chase Jr. \emph{et al.}}}},
  \emph{\bibinfo{title}{JANAF Thermochemical Tables}},
  vol.~\bibinfo{volume}{14} (\bibinfo{publisher}{J.\ Phys.\ Chem.\ Ref.\ Data},
  \bibinfo{year}{1985}), \bibinfo{edition}{3rd} ed., \bibinfo{note}{supplement
  1}.

\bibitem[{REF({\natexlab{a}})}]{REF:STP}
\bibinfo{note}{At STP and saturated H$_2$O the thermolysis of water requires
  $\approx 0.50$~eV, which is within the error of these DFT calculations.}

\bibitem[{REF({\natexlab{b}})}]{REF:Holes}
\bibinfo{note}{Calculated by adding an extra half electron per surface unit
  cell then subtracting the normal density.}

\bibitem[{\citenamefont{Donnerberg et~al.}(1997)\citenamefont{Donnerberg,
  T\"{o}bben, and Birkholz}}]{Donnerberg}
\bibinfo{author}{\bibfnamefont{H.}~\bibnamefont{Donnerberg}},
  \bibinfo{author}{\bibfnamefont{S.}~\bibnamefont{T\"{o}bben}},
  \bibnamefont{and} \bibinfo{author}{\bibfnamefont{A.}~\bibnamefont{Birkholz}},
  \bibinfo{journal}{J.\ Phys.: Condens.\ Matter} \textbf{\bibinfo{volume}{9}},
  \bibinfo{pages}{6359} (\bibinfo{year}{1997}).

\bibitem[{\citenamefont{Bader}(1990)}]{bader}
\bibinfo{author}{\bibfnamefont{R.~F.~W.} \bibnamefont{Bader}},
  \emph{\bibinfo{title}{Atoms in Molecules: a quantum theory}}
  (\bibinfo{publisher}{Clarendon Press}, \bibinfo{address}{Oxford},
  \bibinfo{year}{1990}).

\end{thebibliography}
\end{document}